\documentclass[fleqn,oneside]{article}
\usepackage{epsf,multicol}
\usepackage[T2A]{fontenc}
\usepackage[english]{babel}
\usepackage{amstext}
\usepackage{amssymb}
\usepackage{cite}
\usepackage{graphicx}
\title{A LATTICE MODEL OF INTERCALATION}
\author{T.S. MYSAKOVYCH, V.O. KRASNOV, I.V. STASYUK  \thanks
{Institute for Condensed Matter Physics, Ukraine, 
 Lviv; email (T.S. Mysakovych): mysakovych@icmp.lviv.ua}}
\begin{document}
\maketitle
\begin{multicols}{2}
{\small {The thermodynamics of the lattice model of intercalation of
ions in crystals is considered in the mean field approximation.
Pseudospin formalism is used for the description of interaction of
electrons with ions and the possibility of hopping of intercalated
ions between different positions is taken into account. Phase
diagrams are built. It is shown that the effective interaction between intercalated ions can lead to phase separation or to appearance of modulated phase (it depends on filling of the electron energy band). At high values of the parameter of  ion transfer the ionic subsystem can pass to the superfluid-like state.}}
%\keywords intercalation, phase transition, pseudospin-electron model
%\pacs 64.75.+g, 71.20.Tx, 64.70.-p

\section{Introduction}

Theoretical investigation of intercalation of ions in crystals
 is an actual problem of modern physics. Metal 
oxides as hosts for ion (for example, lithium ions) insertion are very promising electrode materials.
It should be noted that theoretical descriptions of such processes in  most cases
were restricted  to the numerical ab-initio and
density-functional calculations. For example,
 in \cite{stashans,koudriach_2001,koudriach_2002} quantum-chemical Hartree-Fock and
 density-functional calculations were performed to investigate lithium intercalation in TiO$_{2}$ crystal.
 It was shown that Li is almost fully ionized once intercalated (Li looses its valence electron) 
 and reconstruction of electron spectrum at intercalation takes place.
 Thus, ion-electron interaction can play a significant role.
Another interesting feature of  such  crystals  is a 
 shift 
 of  the chemical potential at intercalation 
 into 
the conduction band. As a result, these crystals have metallic conductivity (\cite{ceder}, for a review, see also \cite{modul}); before  intercalation, such crystals are semiconductors with wide gap.
 At intercalation of  lithium in  TiO$_{2}$, phase separation into
Li-poor (Li$_{\sim 0.01}$TiO$_{2}$) and Li-rich (Li$_{\sim 0.5-0.6}$TiO$_{2}$) phases occurs. 
This two-phase
behaviour leads to a constant value of electrochemical potential \cite{wagem_2001,wagem_2003} (this fact is used when
 constructing   batteries). 
In \cite{kyu} the Monte Carlo simulation was performed to investigate the
intercalation using the Hamiltonian which included the interaction between ions only. 

In our previous works \cite{my_jps2007,last2008} we have formulated the pseudospin-electron model of intercalation
and have taken into account the ion-electron interaction. It has been revealed that the effective
 attractive interaction between ions was formed  and the condition
 of appearance of phase separation  has  been established.   The ion-electron interaction 
 was also considered  in \cite{dubl} at the investigation of thermodynamics of 
 $S=1$
 model of intercalation (the model was similar to the known Blume-Emery-Griffiths model), but the electron as well as ion transfer was not taken into account. It should be noted that 
models of pseudospin-electron model type  are widely used in physics of the strongly-correlated electron systems in recent
years. Application of this model to high-temperature superconductors allows one to describe
thermodynamics of anharmonic oxygen ion subsystem and  explain the appearance of inhomogeneous states and the
bistability phenomena (\cite{High}). Such a  model can also be applied to the description of hydrogen-bonded systems. 

In the present paper we deal with a  more complicated   model and 
take into account the possibility of the transfer of intercalated
 ions.
% This model is similar to that used for the description of a system of coexisting itinerant electrons and %local pairs when the
% creation and destruction
%  operators for local pairs (hard-core bosons) obey the Pauli spin 1/2 commutation rules (for example, see %\cite{robasz2004}). But, it should be noted that in
%\cite{robasz2004} the chemical potential of the local pairs and itinerant electrons was the same and the %regime
%of the fixed total number of particles was used. 
 The considered model corresponds to the hard-core boson approach.
 Hard-core bosons obey the Pauli spin-1/2 commutation rules. Since the original  work of Mahan \cite{mahan}, 
 such models were applied for description of ionic conductors and  calculation of their conductivity. Recently   one-particle spectrum was investigated in one-dimensional limit \cite{dulepa}.
A system of hard-core bosons is a
particular case of the well-known Bose-Hubbard model, which has been intensively investigated in the last 15 years (see, for
example, \cite{hcbozons}). The model is  of great interest  due to the  experimental realization of optical lattices (see, for instance, \cite{greiner}).  This model can be directly applied to investigate 
 such objects. The Hamiltonian of the Bose-Hubbard model includes two terms, one is connected with the on-site interaction $U$ between particles and another term is connected with the particle hopping between sites (particles in this model obey the bosonic commutation rules). In the limit $U\rightarrow \infty$ this model reduces to the hard-core boson model. Different theoretical methods were used to study this model: mean-field theory \cite{krish}, random phase approximation \cite{konabe,matsumoto}, strong coupling approach \cite{freericks}, quantum Monte-Carlo method \cite{monte}. Recently a bosonic version of dynamical mean field theory was formulated \cite{bychzuk}. The existence of superfluid and Mott-insulator phases is a characteristic feature of this model.

In addition to our previous investigations \cite{my_jps2007}, the aim of this  work  is the 
study of the ion transfer influence on equilibrium states of intercalated ion subsystem. As it was shown in 
 \cite{my_jps2007,last2008}, the effective interaction between ions 
 can  change (depending on electron band filling) its character from repulsion to attraction, 
 leading to the
 charge-ordered modulated phases or phase separation into uniform phases with different particle concentrations, respectively. Ion hopping between local positions is unfavorable for the realisation 
 of such phases or phase transitions. Besides,  ion hopping leads to the appearance of superfluid type phase. 

We investigate phase transitions in the intercalated ion subsystem within the framework of the lattice model 
 with ion transfer in the regime of the fixed chemical potential of the ions and electrons. Electron subsystem is described by  partially filled one energy band.

\section{The model}

 We consider the following model Hamiltonian:
\[H=\sum_{ij} \Omega_{ij} S^+_i S^-_j \!+\! \sum_{ij\sigma} t_{ij} c^+_{i\sigma} c_{j\sigma} \!+\! \sum_{i\sigma} (g
S^z_i n_{i\sigma}\!-\!\mu n_{i\sigma})- \]
\begin{equation}-\sum_i hS^z_i.\end{equation}
 Here we introduce the pseudospin variable $S_i^z$ which takes two values; $S_i^z=1/2$ when there is an 
 intercalated ion
   in a site $i$  and $S_i^z=-1/2$ when there is no  ion,
$c^+_{i\sigma}$ and $c_{i\sigma}$ are electron creation and annihilation operators,
      respectively. We take into account the possibility of ion and electron jumps between
sites (the first and the second term in (1)) and
       interaction of electrons with ions ($g$ term). The last one is connected with the electron band shift at intercalation (such an effect is known, for example, for the system Li$_x$TiO$_2$ 
\cite{koudriach_2002}); 
$\mu$ and $h$ play
  the role of the chemical potentials of electrons and ions.

It should be noted that we do not consider here the direct interaction between ions. In our 
previous paper
 \cite{my_jps2007},
it was shown that  ion-electron interaction leads to the formation of the effective interaction beweeen ions and even at repulsive direct ion-ion interaction the effective ion-ion interaction of attractive type can be formed. This  can lead to the phase transition of the first order
 between uniform phases with jumps of ion and electron concentrations.

The thermodynamics of the model is investigated in the mean field approximation (MFA)
\[
g n_i S^z_i \rightarrow g \langle n_i \rangle S^z_i +
g n_i \langle  S^z_i \rangle   -
g \langle  n_i \rangle   \langle  S^z_i \rangle \]
\begin{equation}
\Omega S^+_iS^-_j \rightarrow \Omega \langle S^+_i \rangle S^-_j +
\Omega S^+_i \langle  S^-_j \rangle   -
\Omega \langle  S^+_i \rangle   \langle  S^-_j \rangle,
\end{equation}
 here the average ion concentration $w= \langle S^z \rangle +1/2 $ is introduced; in our approximation
 $\langle S^+ \rangle =\langle S^- \rangle =\langle S^x\rangle$, $\langle S^y\rangle =0$.
 The average value $\langle S^x  \rangle$ plays the role of order parameter for the case of the
 superfluid phase (this is a phase with condensate of bose-type), and determines the concentration of condensed particles.

Application of  the MFA to the strongly correlated systems in the limit of a weak one-site  correlation
 makes it possible  to satisfactorily describe their properties, when there is no 
correlational splitting of the electron band. 
This approximation is an analogy to the virtual crystal approximation, which is often used for mixed
 systems. To go beyond the MFA one can use more complicated approximations, for instance, the coherent potential-like approximations. 
Besides, in the case of the Bose-Hubbard model the kinetic energy term is often considered  within the mean field approach. This approximation is well known to give a reasonable estimate of the critical on-site repulsion at which Mott-insulator - superfluid phase transition occurs \cite{hcbozons,krish}.

In \cite{my_jps2001,my_cmp2002} the case $\Omega=0$ was considered. It was shown that 
 if the chemical potential is  near the band center, the double modulation phase is realised in the system, while in the case when the chemical potential is close to the band edges, 
the phase transition between uniform phases occurs.
At intermediate values of the chemical potential the incommensurate modulated phase 
 appears.
 In the present investigation we restrict ourselves to the cases of double modulation phase and uniform phase.

 The Hamiltonian of the model in the
      MFA is as  follows:
\[
H^{MFA}=\sum_{i \alpha \sigma} (g \eta_\alpha - \mu)n_{i\alpha \sigma}+
\sum_{i\alpha}(gn_\alpha-h)S^z_{i\alpha}+ \] 
\[+\sum_{i\alpha,j\beta}t^{\alpha\beta}_{ij}
c^+_{i\alpha\sigma} c_{j\beta\sigma}+\sum_{\alpha\beta i}2\Omega^{\alpha\beta}\langle S^x_\alpha \rangle
S^x_{i\beta}-g\sum_{i\alpha}n_\alpha \eta_\alpha-\]
\begin{equation}
-N\Omega\langle S^x_1 \rangle \langle S^x_2 \rangle,\end{equation}
 here we consider two sublattices:
     $\langle \sum_\sigma  n_{i\alpha\sigma}  \rangle=n_\alpha, \ \
     \langle S^z_{i\alpha} \rangle=\eta_\alpha $;
 $\alpha=1,2$ is a sublattice index, $i$ is an unit cell index,
 $N$ is the number of  lattice sites,
$\Omega\equiv \Omega^{12}=\Omega^{21}=\sum_i \Omega^{12}_{ij}$;
 $\Omega^{\alpha\alpha}=t^{\alpha\alpha}=0$. 

This  Hamiltonian can be diagonalized. We pass to $\bf {k}$-representation and
perform the unitary transformation in the pseudospin subspace
\[
H^{MFA}=\sum_{\alpha\sigma \bf{k}}(\lambda_{\bf{k}\alpha}-\mu) \tilde{n}_{{\bf k}\alpha\sigma}-
\sum_{i\alpha}\tilde{\lambda}_\alpha \sigma^z_{i\alpha}-\]
\begin{equation}
 -g\frac{N}{2}(n_1\eta_1\!+\!n_2\eta_2)
-N\Omega \langle S^x_1 \rangle \langle S^x_2 \rangle \end{equation}
\[ \lambda_{\bf{k}\alpha}=g\frac{\eta _1+\eta
_2}{2}+(-1)^{\alpha} \sqrt{(g\frac{\eta _1-\eta _2}{2})^2+t^{2}_{\bf{k}}}\] \[c_{{\bf{k}} 1\sigma}=
\tilde{c}_{{\bf k} 1\sigma} \cos\phi + \tilde{c}_{{\bf k}2\sigma} \sin \phi, \]
\[c_{{\bf{k}} 2\sigma}= - \tilde{c}_{{\bf{k}} 1 \sigma} \sin\phi +\tilde{c}_{{\bf{k}} 2 \sigma} \cos\phi,\]
\[\sin2\phi=\frac{t_{\bf{k}}}{\sqrt{(g\frac{\eta _1-\eta _2}{2})^2+
t^{2}_{\bf{k}}}}\] \[S^z_{i \alpha}=\sigma^z_{i\alpha}\cos\theta_\alpha+\sigma^x_{i\alpha} \sin\theta_\alpha,\]
\[S^x_{i \alpha}=\sigma^x_{i\alpha}\cos\theta_\alpha-\sigma^z_{i\alpha}
\sin\theta_\alpha,\ \ \ \sin \theta_\alpha= \frac{2\Omega \langle S^x_\beta \rangle}{\tilde{\lambda}_\alpha},\]
\[\tilde{\lambda}_{\alpha}=\sqrt{(gn_{\alpha}-h)^2+(2 \Omega \langle S^x_\beta \rangle )^2},\ \ \
\alpha\neq\beta.\]

 The doubling of unit cell  leads to the splitting in the electron spectrum.
 Two subbands are separated by the gap $g|\eta_1-\eta_2|$.
The electron  band  changes its  position at intercalation.

  Using MFA, we can calculate the mean values of both the electron and ion concentrations:
\[
n_{\alpha}{=}\frac{1}{N}\sum_{\bf{k} \sigma}(\frac{1+\cos 2 \phi}{2} (e^{\frac{\lambda_{\bf{k}
\alpha}-\mu}{T}}+1)^{-1}+ \]
\begin{equation}\label{equation}
+\frac{1-\cos 2 \phi}{2} (e^{\frac{\lambda_{\bf{k}
\beta}-\mu}{T}}+1)^{-1}),\end{equation}
\[\eta _{\alpha}=\frac{h-gn_{\alpha}}{2 \tilde{\lambda}_{\alpha}}
\tanh(\frac{\beta \tilde{\lambda}_{\alpha}}{2}),
\langle S^x_\alpha \rangle=-\frac{2\Omega \langle S^x_\beta
\rangle  } {2\tilde{\lambda}_\alpha}\tanh(\frac{\beta \tilde{\lambda}_{\alpha}}{2}).\]

 To find the thermodynamically stable states we also have to calculate the grand canonical potential
\[
\frac{\Phi}{\frac{N}{2}}=-\frac{T}{N}\sum_{\bf{k}
\sigma}\ln((e^{\frac{\mu-\lambda_{\bf{k}1}}{T}}+1)^{-1})(e^{\frac{\mu-\lambda_{\bf{k}1}}{T}}+1)^{-1}))-
\]
\[-T\ln(4\cosh(\frac{\beta \tilde{\lambda}_{1}}{2}\cosh(\frac{\beta \tilde{\lambda}_{2}}{2}))-
g(n_{1}\eta_{1}+n_{2}\eta_{2})-\]
\begin{equation}\label{grand}
-2\Omega \langle S^x_1 \rangle \langle S^x_2 \rangle.
\end{equation}
 The absolute minima of the $\Phi$-function determine the equilibrium states.

 \section{Results}

 The semielliptic density of states,
  $\rho(\epsilon)=\frac{2}{\pi W^2}\sqrt{W^2-\epsilon^2}$,
$-W<\epsilon<W$, where $W$ is a half width of the electron band,  was used 
 ($W$ is chosen as  energy unit; in our calculations  we put $W=1$). Using this density of states,
 we perform summation over $\bf k$ in the equations of self-consistency (\ref{equation}) and in the expression for the grand canonical potential
 (\ref{grand}).

As  noted above, the stable states are 
 obtained using  the condition of minimum of the function 
 $\Phi$.
 In Fig.1 and Fig.2 the $(h-\Omega)$ phase diagrams are shown for the cases $\mu=0$
 (at the center of the band) and $\mu=-0.7W$ (near the lower band edge).

 \begin{center}
\noindent \epsfxsize=0.8\columnwidth \epsffile{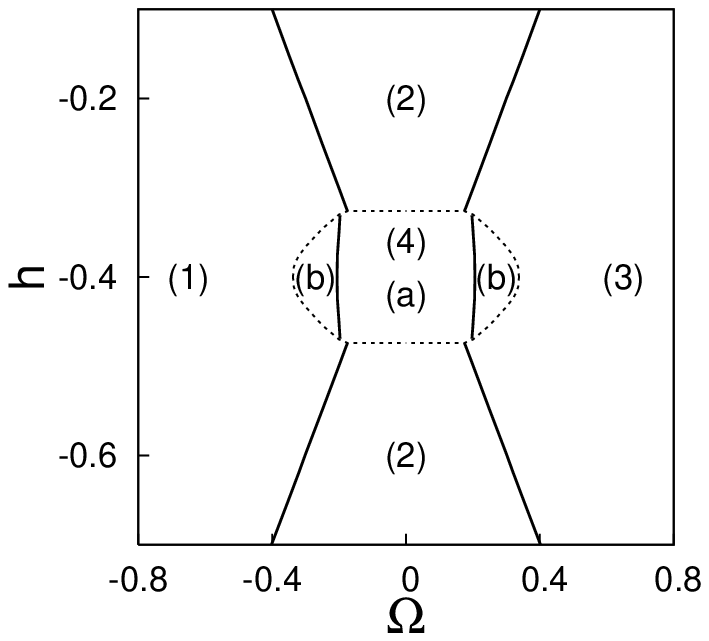}
\end{center}

\vskip-3mm\noindent{\footnotesize Fig.1 Phase diagram in the ($h-\Omega$) plane at $\mu=0$. The parameter values
are: $g=-0.4, W=1, T=0.03$. 
(1),(3) - uniform phase with  $\langle S^x \rangle \neq 0$  ((1): $\langle S^x_1 \rangle=\langle S^x_2 \rangle$,
(3): $\langle S^x_1 \rangle=-\langle S^x_2 \rangle$); 
(2)-uniform phase with $\langle S^x \rangle=0$;
 (4)-modulated phase ((a): $\langle S^x_{1,2} \rangle=0$,
 (b): $\langle S^x_{1,2} \rangle \neq 0$).
 The solid line denotes the phase transitions of the second order and 
 the dotted line denotes the phase transition of the first
 order.}%
\vskip15pt

 In the case $\mu=0$ and at small values of ion hopping parameter $\Omega$ the system undergoes the phase transition of the first order from the uniform to modulated phase
 (in the modulated phase $n_1\neq n_2, \eta_1\neq\eta_2$)
 at the change of the chemical potential of the ions (dotted line in Fig.1). In the case $\mu=-0.7W$ 
 and at small values of $\Omega$ the phase transition of the first order between two uniform phases with jumps of the average ion and electron concentration  (and the phase separation in the regime 
 of the fixed concentration,
 see \cite{my_jps2007,last2008,my_jps2001,my_cmp2002} for more details) takes plase (dashed line in Fig.2).

  It is easy to see  that at high values of the parameter of ion transfer $\Omega$ the only
 possible phases are the uniform phases with $\langle S^x \rangle \neq 0$ and $\langle S^x \rangle = 0$.
 The phase with $\langle S^x \rangle\neq 0$ appears due to the presence of ion
 hopping between sites; this  phase is an analogy to a superfluid phase in the systems of hard-core bosons and can correspond to the state with high mobility 	of intercalated ions.
 In the case $\Omega<0$, one finds  $\langle S^x_1 \rangle = \langle S^x_2 \rangle $,  
while in the case  $\Omega>0$, one finds
 $\langle S^x_1 \rangle =- \langle S^x_2 \rangle $.

\begin{center}
\noindent \epsfxsize=0.8\columnwidth \epsffile{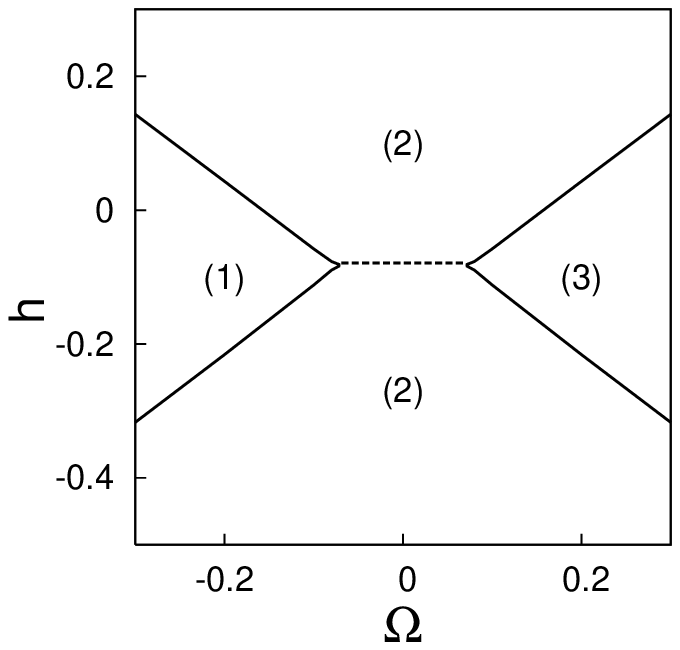}
\end{center}

\vskip-3mm\noindent{\footnotesize Fig.2 Phase diagram in the ($h-\Omega$) plane at $\mu=-0.7$. The parameter values
are: $g=-0.4, W=1, T=0.03$. The notations are the same as in Fig.1.
 The solid line denotes the phase transition of the second order and 
 the dashed line denotes the phase transition of the first order.}%
\vskip15pt

 In Fig.3 and Fig.4 the $(h-\mu)$ phase diagrams are shown for the cases $\Omega=0$ and $\Omega=0.25$. Dotted line denotes 
 the first order phase transition between uniform and modulated phases, dashed line denotes the first order phase transition
 between uniform phases.

\begin{center}
\noindent \epsfxsize=0.8\columnwidth \epsffile{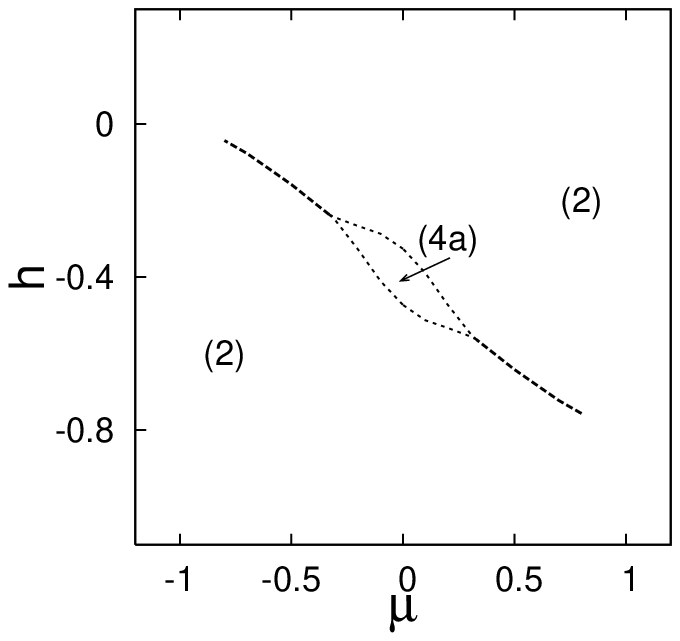}
\end{center}

\vskip-3mm\noindent{\footnotesize Fig.3 Phase diagram in the ($h-\mu$) plane at $\Omega=0$. The parameter values
are: $g=-0.4, W=1, T=0.03$. The notations are the same as in Figs.1,2.}%
\vskip15pt
\begin{center}
\noindent \epsfxsize=0.8\columnwidth \epsffile{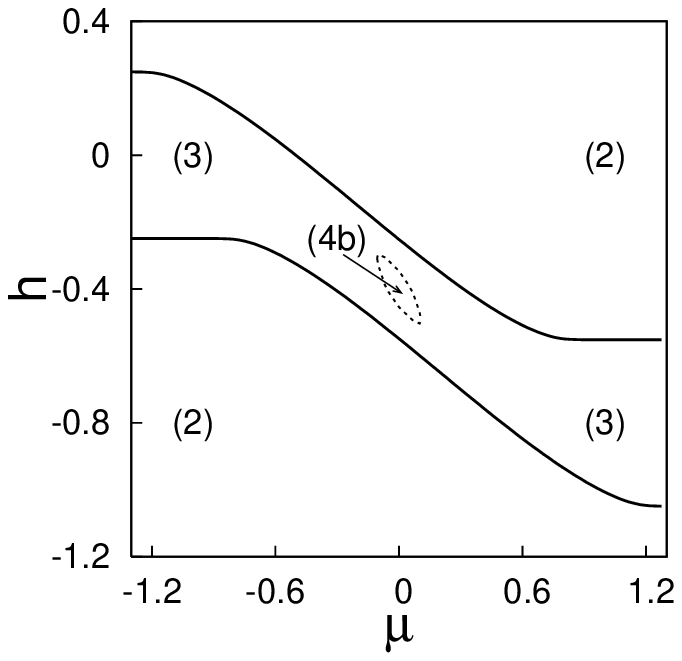}
\end{center}

\vskip-3mm\noindent{\footnotesize Fig.4 Phase diagram in the ($h-\mu$) plane at $\Omega=0.25$. The parameter values
are: $g=-0.4, W=1, T=0.03$. The notations are the same as in Figs.1,2.}%
\vskip15pt

The next four diagrams, which are shown in Figs. 5-8 are the $(T-h)$ phase diagrams.
 In Fig.5 the first order phase transition curve between two uniform phases with jumps
 of ion and electron concentrations is shown. This curve ends in critical point at some 
 value of temperature $T_{cr}$. At high values of $\Omega$ this phase transition disappears (see Figs.2,6).
 The existence of such phase transitions (in the regime of the fixed concentration it corresponds to the phase separation into phases with different concentration of ions) is in accordance with experimental data for intercalated crystals, where the appearance of poor and rich ion concentration phases was observed (see, for example, \cite{wagem_2001}). The presence of modulated phase in intercalated crystals  is also indicated in experiments (for review, see \cite{modul}).

 The phase transition from the 
uniform to modulated phase can be of the second or the first order, this is illustrated in 
 Figs.7,8 for the cases $\Omega=0$ and $\Omega=0.3$. We should draw attention to the fact,
that with increasing temperature  the first order transition will transform into the second order 
 one
 and then will disappear (as it is shown in Figs.7,8).

\begin{center}
\noindent \epsfxsize=0.8\columnwidth \epsffile{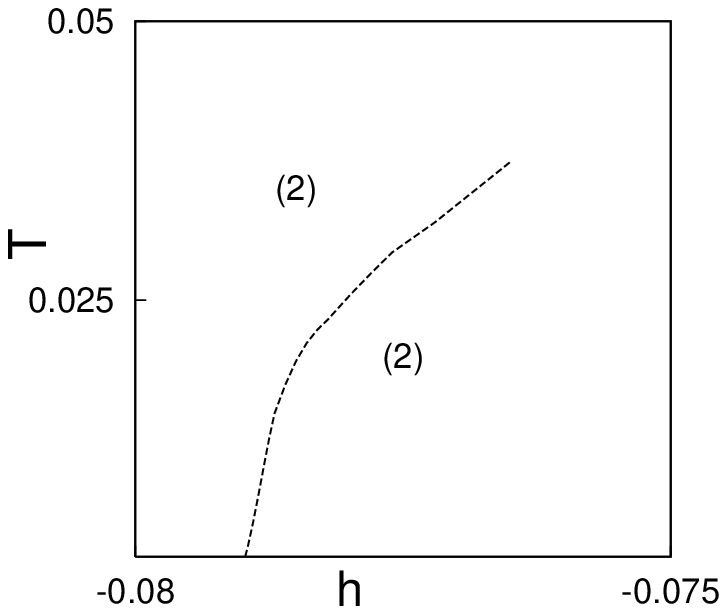}
\end{center}

\vskip-3mm\noindent{\footnotesize Fig.5 Phase diagram in the ($T-h$)
plane at $\Omega=0$. The parameter values are: $g=-0.4, W=1, \mu=-0.7 $. The notations are the same as in Figs.1,2.}%
\vskip15pt

\begin{center}
\noindent \epsfxsize=0.8\columnwidth \epsffile{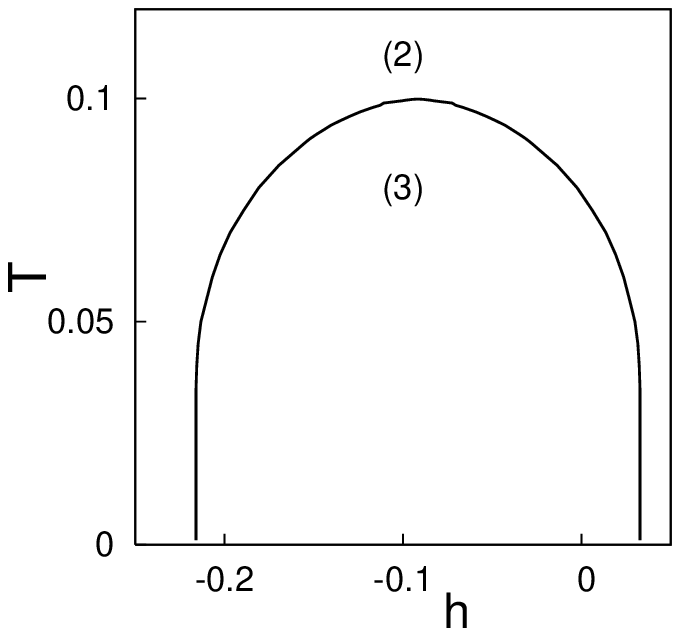}
\end{center}

\vskip-3mm\noindent{\footnotesize Fig.6 Phase diagram in the ($T-h$) plane at $\Omega=0.2$. The parameter values
are: $g=-0.4, W=1,
\mu=-0.7$. The notations are the same as in Figs.1,2.}%
\vskip15pt

\begin{center}
\noindent \epsfxsize=0.8\columnwidth \epsffile{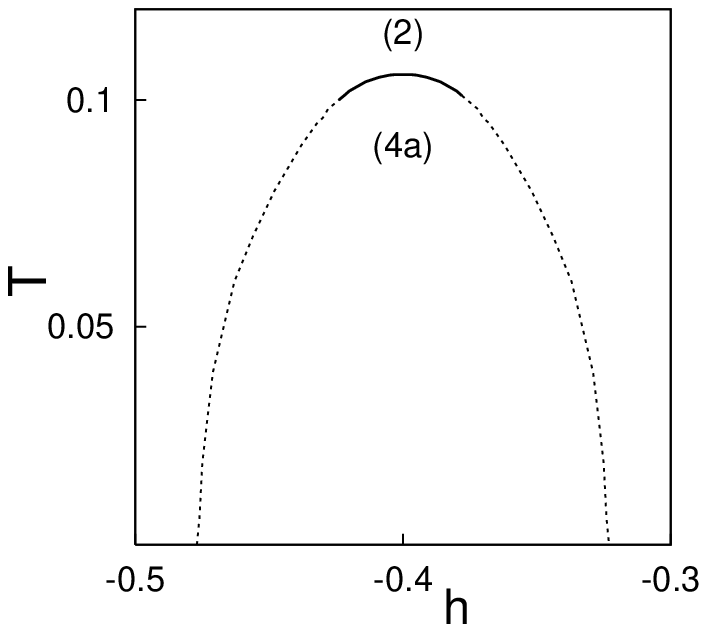}
\end{center}

\vskip-3mm\noindent{\footnotesize Fig.7 Phase diagram in the ($T-h$) plane at $\Omega=0$. The parameter values
are: $g=-0.4, W=1, \mu=0$.
The notations are the same as in Figs.1,2.
}
\vskip15pt

\begin{center}
\noindent \epsfxsize=0.8\columnwidth \epsffile{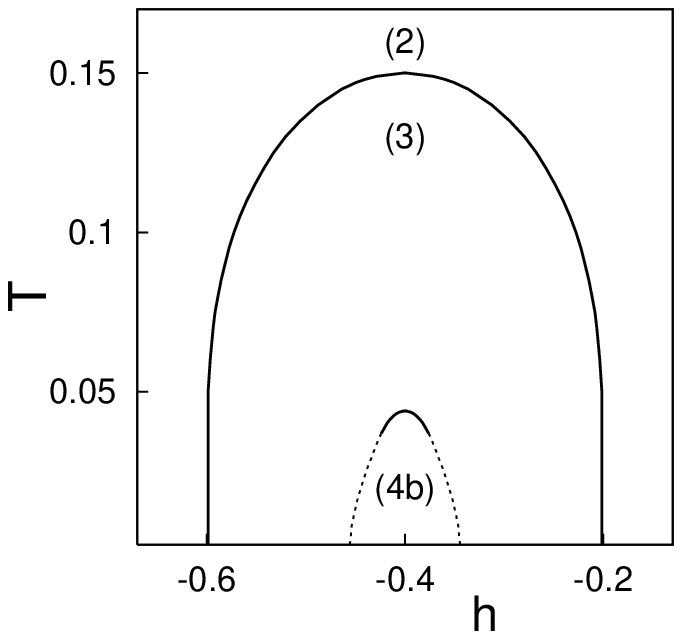}
\end{center}
\vskip-3mm\noindent{\footnotesize Fig.8 Phase diagram in the ($T-h$) plane at $\Omega=0.3$. The parameter values
are: $g=-0.4, W=1, \mu=0$.
The notations are the same as in Figs.1,2.}%
\vskip15pt

 \section{Conclusions}

 In this work, the pseudospin-electron model of ion intercalation in crystals
  has been formulated.
 The model can be applied for the description of the thermodynamics of such a process in materials with electron bands where the band filling has metallic or semimetallic character (in particular,  compounds of transition metals such as TiO$_2$  or other similar systems with narrow conduction bands).
 The thermodynamics of the model has been investigated in the mean-field
   approximation. The effective interaction between intercalated ions is formed due to
 their
 interaction with electron subsystem.
 Such an interaction is attractive or repulsive  depending on the filling of electron band (in the first case the chemical potential of electrons should be close to the band edge; the second case is realised near half filling). 
   The appearance of modulated phase or  phase transitions  of the first order with jumps of ion and electron
 concentrations (in the regime of the fixed concentrations it corresponds to the phase separation) has been
 established.

 Increase of the ion transfer parameter  leads to the disappearance of both of modulated phase and phase transition with jumps of the ion and electron concentrations.
 Besides, the new phase with $\langle S^x \rangle\neq 0$ appears due to  ion
 hopping between sites; this  phase is an analogy to superfluid phase in the systems of hard-core bosons or  superionic phase in the crystalline ionic conductors (phase with high mobility of ions). Such a phase can exist at intermediate values of the chemical potential of intercalated ions and the transition to this phase is  of the second order.
 To investigate this phase in detail we should examine the behaviour of the conductivity and other characteristics of the system.
 This is the task for future investigations.

\end{multicols}
\end{document}